\documentclass[aps,pra,twocolumn,showpacs]{revtex4-1}

\newcommand{\Jeff}{J_\mathrm{eff}}
\newcommand{\be}{\begin{equation}}
\newcommand{\ee}{\end{equation}}

\renewcommand{\a}{\hat{a}}
\newcommand{\n}{\hat{n}}

\usepackage{graphicx}
\usepackage{amssymb,amsfonts,amsmath}
\usepackage[colorlinks=true,citecolor=Cerulean,linkcolor=RubineRed,urlcolor=Cerulean]{hyperref}
\usepackage{graphicx}
\usepackage{color}
\usepackage[usenames,dvipsnames]{xcolor}

\renewcommand{\vec}[1]{\mathbf{#1}}

\newcommand*{\affaddr}[1]{\emph{#1}}
\newcommand*{\affmark}[1][*]{\textsuperscript{#1}}

\begin{document}

\title{Realization of uniform synthetic magnetic fields by periodically
shaking an optical square lattice}

\author{C.E.~Creffield\affmark[1], G.~Pieplow\affmark[1], F.~Sols\affmark[1,2,3], and N.~Goldman\affmark[4]\\
	\affaddr{\affmark[1]Departamento de F\'isica de Materiales, Universidad
		Complutense de Madrid, E-28040 Madrid, Spain}\\
	\affaddr{\affmark[2]ICFO-Institut de Ci\`encies Fot\`oniques, The Barcelona Institute of
		Science and Technology, E-08860 Castelldefels (Barcelona), Spain}\\
	\affaddr{\affmark[3]Department of Physics, Harvard University, Cambridge, MA 02138, USA}\\
	\affaddr{\affmark[4]CENOLI, Facult{\'e} des Sciences, Universit{\'e} Libre de Bruxelles (U.L.B.), B-1050 Brussels, Belgium}
}

\date{\today}


\begin{abstract}
Shaking a lattice system, by modulating the location of its sites periodically in time, is a powerful method to create effective magnetic fields in engineered quantum systems, such as cold gases trapped in optical lattices. However, such schemes are typically associated with space-dependent effective masses (tunneling amplitudes) and non-uniform flux patterns. In this work we investigate this phenomenon theoretically, by computing the effective Hamiltonians and quasienergy spectra associated with several kinds of lattice-shaking protocols. A detailed comparison with a method based on moving lattices, which are added on top of a main static optical lattice, is provided. This study allows the identification of novel shaking schemes, which simultaneously provide uniform effective mass and magnetic flux, with direct implications for cold-atom experiments and photonics.
\end{abstract}

\maketitle

\section{Introduction}

Using one quantum system to simulate another, an idea popularized by Feynman \cite{feynman1982}, is a fascinating and rapidly developing topic of current research~\cite{Cirac2012,Georgescu2014}. Hamiltonians arising from many different areas such as
condensed matter and high-energy physics
can be hard to study computationally or in the laboratory, because they require
resources  or parameter regimes that are difficult or impossible to attain.
Quantum simulators offer an attractive means to circumvent such difficulties.
Many different physical platforms have been proposed as quantum simulators,
including ultracold gases \cite{Lewenstein2012,Bloch2008},
trapped ions \cite{Blatt2012}, superconducting circuits
\cite{Houck2012}, and photonics \cite{Aspuru2012}.

One particular condensed matter problem that can be simulated with 
engineered quantum systems, and which constitutes the core of the present paper, is the spectrum of electrons moving on a lattice subjected to a uniform magnetic field. This forms intriguing fractal structures known as ``Hofstadter butterflies''~\cite{hofstadter,avron}, which are only visible in regimes of extremely large magnetic flux densities, unreachable in conventional solid state systems (see also~\cite{Dean2013}). In a more general context, we may note that, in recent years, more and more effort has been directed at proposing and performing experiments aimed at realizing synthetic gauge fields (e.g.~artificial
magnetic fields, spin-orbit coupling) and topological phases, in a wide range of physical systems.
Some examples include light in photonic lattices \cite{fang, hafezi,rechtsman2013,ling2014,dubvcek2015}, phonons in ion traps \cite{yang2016, bermudez2012}, microwave networks \cite{hu2015}, sound and light in cavity optomechanics \cite{peano2015},
mechanical systems \cite{mechanical1,mechanical2},
and atoms in optical lattices \cite{Jaksch,Gerbier,DalibardReview,kolovsky,struck2012,struck2013,hauke2012,ketterle, bloch, GoldmanReview,goldman2014, HaldaneETH,aidelsburger2015}.
Schemes for simulating artificial magnetic fields are generally based on modifying
the system's hopping terms, so that they become complex-valued, thus acquiring phase-factors
that correspond to Aharonov-Bohm phases~\cite{GoldmanReview}.
A powerful means of achieving this has become known as
``Floquet engineering''~\cite{Kitagawa,goldman2014,FloquetReview,bukov2015}, in which a rapidly oscillating field is used
to manipulate the properties of the system by producing an effective
(time-independent) Hamiltonian with the desired properties. In such schemes the energy spectrum
of the original undriven system is replaced by the Floquet spectrum of quasienergies
of the driven model, which amounts to the energy spectrum of the effective time-independent model~\cite{Kitagawa,goldman2014,FloquetReview,bukov2015}.

Shaking the lattice at a high frequency,
i.e.\ rapidly oscillating the position of the lattice sites,
is the method we focus on in this work to modify hopping terms and generate effective magnetic fields.
This method is extremely general, and can be applied to a wide range of lattice
systems, including cold atoms in optical lattices and arrays of photonic waveguides.
Indeed shaking is one of the few experimental tools available
to manipulate the tunneling terms in photonic crystals \cite{rechtsman2013,mukherjee2015,mukherjee2016}.
For cold atom systems, other schemes such as placing ``moving lattices'' on top of
the underlying optical lattice potential \cite{ketterle, bloch,aidelsburger2015}, or using the internal structure of the atoms to generate synthetic gauge fields \cite{Jaksch,Gerbier,celi2014}, are
available. Shaking optical lattices~\cite{kolovsky,hauke2012,struck2012,struck2013,HaldaneETH}, however, is one of the most simple techniques since it generally does not require additional lasers, and beyond the coupling to the lattice
potential, the internal (hyperfine) structure of the atoms is not important.
It is therefore crucial to investigate general shaking schemes, to
determine their advantages and weaknesses, and assess their ability to
create uniform stable fluxes in the various physical platforms where
shaking is generally available.

In this paper, we consider schemes based on {\em resonant} shaking
\cite{kolovsky, bermudez2012,hauke2012,ketterle, bloch,aidelsburger2015,Creffield:2014} to generate
homogeneous magnetic fields in a square lattice
by suitably modifying the tunneling terms.
In such schemes the inertial force associated with
the shaking produces a potential that contains both a static
and an oscillating component, and the term ``resonant'' refers to the matching of
the static part with the oscillation frequency.
Simple resonant shaking of the lattice \cite{kolovsky,bermudez} can be used to
produce a uniform magnetic flux, but has the disadvantage that the effective
mass is spatially dependent \cite{comment}. A development of this
scheme, termed ``split-driving'' \cite{Creffield:2014}, solves the
problem of the mass inhomogeneity, but at the cost of rendering the
flux weakly space-dependent. We show how considering the origin of
these two effects allows us to design different shaking schemes which 
makes it possible to avoid both these problems and achieve the ideal result:
a uniform artificial magnetic field,
in which the particles' effective mass is homogeneous. We benchmark the
various schemes against each other, and show how they provide a powerful
and convenient means to produce artificial gauge fields in lattice systems.

\subsection{Outline}
The paper is structured as follows. In Section \ref{peierls_section}
we briefly discuss the tight-binding description of electrons in a square lattice under the influence of an external magnetic field. This provides the connection between complex hopping elements and magnetic fields.
In Section \ref{accel_section} we then comment on the appearance of inertial forces in the
tight binding description and provide some context for some commonly used unitary
transformations that connect the Hamiltonian in the rest frame of the lattice to
a Hamiltonian where the shaking and linear force terms solely enter in the complex
hopping matrix elements. We then proceed to investigate several shaking schemes.
In Section \ref{sine_driving}, the most simple scheme, sinusoidal driving, provides
an introduction to the effective Floquet Hamiltonian in the high frequency regime.
In Section \ref{2_step} we introduce the two-step split-driving scheme, which produces
a uniform effective mass, but a weakly-varying flux pattern, and in
Section \ref{4_step} we show how changing the split-driving to a four-step scheme
succeeds in correcting both problems, to produce a uniform magnetic
field with a constant effective mass.
We then proceed in Section \ref{waveforms} to show how changing the
shaking from a sinusoidal to other waveforms can be used to
minimize the impact of the inhomogeneity in the flux of the two-step scheme. In Section \ref{compare}, we make a quantitative comparison between the
different schemes, specifically, shaken vs. moving lattice approaches. Finally we give our conclusions and outlook in Section \ref{concs}.

\section{Peierls phase factors and flux per plaquette \label{peierls_section}}

\begin{figure}[t!]
\begin{center}
\includegraphics[width = .35\textwidth,clip=true]{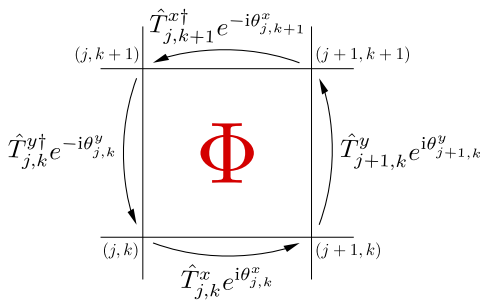}
\end{center}
\caption{The flux per plaquette is defined by the accumulated phase of a particle that loops around a plaquette of the lattice.}
\label{fig:flux_per_plaq}
\end{figure}

In order to introduce notation and provide some context
for how a synthetic magnetic field can be simulated in engineered quantum systems, we start by providing a brief overview of how
a particle experiences a magnetic field on a lattice.
We consider a typical two-dimensional optical square lattice, formed by
the superposition of two optical standing waves.  Generalization to other physical platforms (e.g.~photonic crystals) and geometries is straightforward.
When the optical
lattice potential is sufficiently deep, and interactions between
the atoms are weak, the dynamics of cold atoms
moving in the lattice can be well described by a single-band
tight-binding model
\begin{equation}
\hat{H}_0 = -J \sum_{j, k} (
\a_{j,k+1}^\dagger \a_{j,k} +
\a_{j+1,k}^\dagger \a_{j,k}) + {\rm H.c.}
\label{tight_binding}
\end{equation}
where $J$ is the tunneling amplitude between neighboring sites.
In quantum mechanics, the influence of a magnetic field $\bf B\!=\! \bf{\nabla} \!\times\! \bf{A}$ on the behavior
of a charged particle moving in a continuum is described by modifying the canonical
momentum with the vector potential, ${\bf p} \rightarrow {\bf p} - e \bf{A}(\bf{r})$, where $e$ denotes the elementary charge.
When the system is defined on a {\em lattice}, the momentum operator
is replaced by hopping operators connecting neighboring sites,
and the vector potential enters in the form of phase factors \cite{hofstadter}, termed
Peierls phase-factors, which modify the single-particle hopping terms as
\begin{align}
\hat{a}^\dagger_{j+1,k}\hat{a}_{j,k}&=\hat{T}^x_{j,k}\longrightarrow \hat{T}^x_{j,k} \, e^{ {\rm i} \theta_{j,k}^x}~,
\\
\hat{a}^\dagger_{j,k+1}\hat{a}_{j,k}&=\hat{T}_{j,k}^y\longrightarrow \hat{T}_{j,k}^y \, e^{{\rm i} \theta_{j,k}^y}~,
\end{align}
where $\a_{j,k} / \a_{j,k}^\dagger$ are the standard annihilation/creation
operators for an atom (bosonic or fermionic)
on lattice site labeled $(j,k)$, as shown in
Fig.~\ref{fig:flux_per_plaq}. The phases are defined by
the following line integrals
\begin{align}
\begin{aligned}
\label{phase_factors}
&\theta_{j,k}^x =e\int_{\vec{r}_{j,k}}^{\vec{r}_{j+1,k}}\vec{A}(\vec{r},t)\cdot d\vec{x} ~,\\
&\theta_{j,k}^y = e\int_{\vec{r}_{j,k}}^{\vec{r}_{j,k+1}}\vec{A}(\vec{r},t)\cdot d\vec{y}~.
\end{aligned}
\end{align}
Performing the Peierls substitution thus leads to the well-known
Harper-Hofstadter Hamiltonian
\begin{align}
\begin{aligned}
\hat{H}_{\rm H} = -J \sum_{j, k}
\left( \a_{j,k+1}^\dagger \right. \a_{j,k} &e^{ {\rm i} \theta_{j,k}^y}
\\
+
& \left. \a_{j+1,k}^\dagger \a_{j,k}e^{ {\rm i} \theta_{j,k}^x} \right) + {\rm H.c.}
\end{aligned}
\label{tight_binding_harper_hof}
\end{align}
In the continuum case, the magnetic flux passing through an area bounded
by a curve $C$ is given by the line-integral of the vector potential
\begin{equation}
\Phi(C) = \int \vec{B} \cdot d \vec{S} = \oint_C \vec{A} \cdot d \vec{l} \ .
\label{curl_B}
\end{equation}
For the case of a lattice, the quantity of interest is the magnetic
flux passing through a given plaquette. Comparing Eq.~\eqref{curl_B} with
Eq.~\eqref{phase_factors}, this is clearly given by the sum of the Peierls
factors for a particle moving anti-clockwise around the plaquette,
as shown in Fig.~\ref{fig:flux_per_plaq},
\begin{equation}
\Phi(j,k) = \theta^x_{j,k} + \theta^y_{j+1,k} - \theta^x_{j,k+1} - \theta^y_{j,k} \ .
\label{peierls}
\end{equation}

The simulation of a lattice system subjected to a magnetic field thus
amounts to inducing the appropriate phase factors on the hopping terms.
The magnetic flux per plaquette $\Phi(j,k)$ is a gauge invariant quantity.
However, the vector potential, and thus the Peierls factors, are not.
To evaluate the Peierls factors we must first choose a particular gauge.
To generate a uniform magnetic field,
two common examples are the symmetric gauge,
$\vec{A} = (B / 2) (-y , x )$, and the
Landau gauge $\vec{A} = -B (y, 0)$.
The Landau gauge has a particularly simple form as it only
involves generating phases in one direction. Because of this it is the most practical for experimental realization and accordingly it is the
gauge that we consider in the following of this work.

Our aim is thus to identify schemes that modify the tunneling matrix elements along the $x$-direction,  producing $y$-dependent tunneling phases of the form \begin{equation}
\theta^x_{j,k} = \Phi k, \quad \theta^y_{j,k} = 0 ,
\label{landau_gauge}
\end{equation}
which correspond to the Landau gauge  for a lattice system.

%

\section{Accelerated optical lattices \label{accel_section}}

\subsection{Tight-binding approach and change of frame}
We now consider the effect of moving the optical
lattice in the $x$-direction with a time-dependent acceleration
$a(t)$. This can be achieved experimentally by introducing a phase
modulator to the laser beams in the $x$-direction, or if the optical
standing wave is produced by reflection from a mirror, by physically
moving the mirror in space with a piezo-actuator \cite{arimondo,HaldaneETH}. Note that in photonics, such modulations can be directly imprinted using femtosecond laser writing~\cite{rechtsman2013,mukherjee2015}.
If we now transform to the rest-frame of the lattice, this
acceleration gives rise to an inertial force (see Appendix \ref{appendix:frames} for more detail), which can be included in the Hamiltonian as
\begin{equation}
\hat{H}_{\rm latt}(t) = \hat{H}_0 +
\sum_{j,k} \left[ j m a(t) + \frac{m}{2} a(t) d(t) \right] \n_{j,k} \ ,
\label{ham_lattice}
\end{equation}
where $d(t)$ is the spatial displacement of the lattice,
$a(t) \!=\! \ddot d(t)$ is the lattice acceleration, $m$ is the effective mass of the atoms, and
$\hat n_{j,k}$ is the standard number operator.
Throughout the paper we measure distances in units of the lattice spacing,
and thus, for example, $j$ is the $x$-coordinate of
lattice site $(j,k)$.
We note that the final term, is independent of
$j$, and so simply gives rise to an overall phase. Accordingly,
as this gives no physical effect, we can drop this term, leaving
the inertial force described by a potential that grows linearly
along the $x$-direction of the lattice, $V(t) = j m a(t)$.

A shift of the momenta (see Appendix \ref{appendix:frames}) from the
moving to the laboratory frame is given by
a unitary transformation, defined by the operator
\begin{align}
\hat R (t)= \exp \left ( {\rm i} m \sum_{j,k} j v(t) \n_{j,k}  \right ) , \label{movingframe}
\end{align}
where $v(t) \!=\! \dot d(t)$ is the lattice velocity.
Equation (\ref{movingframe}) applied to (\ref{ham_lattice}) removes the linear potential
term and generates a vector potential in the continuum representation, or hopping phases in a tight-binding picture. A detailed description is given in Appendix \ref{appendix:frames} within the continuum representation. The tight-binding case is discussed in the next section.

\subsection{Periodic driving \label{standard_shaking}}
A uniform acceleration, $a(t) = a_0$, thus has the effect of introducing
a static tilt to the lattice [Eq.~\eqref{ham_lattice}].
We shall consider the more general case when in addition to
the static tilt, the acceleration also varies periodically, so as to
``shake'' the lattice.
In the high-frequency limit, i.e.~when the shaking
frequency is the dominant energy scale of the system, its long-time dynamics can be
well captured by a time-independent (effective) Hamiltonian. This effective Hamiltonian typically includes renormalized hopping
terms $J \rightarrow \Jeff$. As we show in Appendix \ref{F_factors},
this renormalization can be calculated explicitly for the general case, using a perturbative expansion
in orders of $1 / \omega$.
For example, when the shaking has a simple (single-harmonic) sinusoidal time-dependence, the
renormalization takes the well-known Bessel function form \cite{bessel,Dunl86,hanggi}.

In the following, we consider shaking the lattice along the $x$ direction, and we shall introduce a further degree
of freedom, namely, that the temporal phase of the shaking varies with the
$y$-coordinate. We shall see later that such a spatial variation is essential
to produce synthetic gauge fields \cite{kolovsky, bermudez},
as otherwise the renormalized hoppings will be uniform, and so the sum
of the Peierls factors around a plaquette (\ref{peierls})
will be identically zero.

We therefore introduce a $y$-dependent temporal phase $\theta_k$, and take the
Hamiltonian to have the form
\begin{equation}
\hat{H}_{\rm latt}(t) =  \hat{H}_0 +
\sum_{j,k} j \left[ V_0 + f(\omega t + \theta_k) \right] \n_{j,k} \ ,
\label{ham_driven}
\end{equation}
where $V_0$ is the static lattice tilt, and
$f(\omega t)$ is a $T$-periodic driving function.
We note that as well as arising from a uniform acceleration of
the lattice, the tilt can also
be generated by other means such as a magnetic \cite{bloch}
or gravitational field gradient \cite{nagerl}, or
approximately by applying a dipole potential \cite{ketterle}.

We note that the momentum-shift operator in Eq.~\eqref{movingframe} is now explicitly given by
\begin{align}
\hat R (t)\!=\!\exp \left ( {\rm i} \sum_{j,k} \left [ V_0 t + F(\omega t + \theta_k)-F(\theta_k) \right ] j \n_{j,k}  \right ) ,
\label{movingframe_bis}
\end{align}
where $F(\omega t) \! :=\! \int^t_0 \ f(\omega t') dt'$ .

\section{Sinusoidal driving \label{sine_driving}}

We first examine the specific choice of a sinusoidal driving, together with a {\em resonant tilt}, such that
\be
f(\omega t) = K \sin \omega t , \quad V_0 = N \omega, \quad N\in \mathbb{Z}.\label{sinusdrive}
\ee
The driven Hamiltonian in Eq.~\eqref{ham_driven} then produces the well-known
phenomenon of photon-assisted-tunneling phenomenon \cite{arimondo_2}, which is associated with the renormalization of the Hamiltonian's tunneling amplitudes
by Bessel functions of the first kind, $\Jeff = J \ {\mathcal J}_N (K_0)$.
For convenience we introduce the dimensionless variable $K_0 = K / \omega$,
and take $\hbar = 1$ throughout this work.

The effective Hamiltonian associated with the sinusoidally-driven lattice has been analyzed previously in the literature \cite{sine_analysis}. However, we will present a derivation of this operator here in some detail, both to illustrate our methods  of
analysis, and to point out the problems in using this form of shaking
to generate synthetic magnetic fields. First, we note that the driven Hamiltonian in Eq.~\eqref{ham_driven}, with Eq.~\eqref{sinusdrive}, contains two diverging terms in the limit $\omega \rightarrow \infty$ (typically $K\!\sim\!\omega$ in experiments); these can be removed by shifting
the momenta into the laboratory frame
\begin{align}
&\vert \Psi (t) \rangle \rightarrow \vert \Psi ' (t) \rangle = \hat R (t) \vert \Psi (t) \rangle ,
\\
&
\begin{aligned}
 &\hat R (t) = \exp \left ( {\rm i} \sum_{j,k} j \right. \n_{j,k}  \Big \{  N \omega t
 \\
& \hspace{2cm}  - K_0[ \cos (\omega t + \theta_k) -\cos(\theta_k)]\Big \}\left.\vphantom{\sum_{j,k}} \right ) , \label{R-trans-TB}
\end{aligned}
\\
& \hat{\mathcal{H}}(t) = \hat{R}(t) \hat{H}_\mathrm{latt}(t) \hat R^\dagger(t) +
{\rm i}\frac{{\rm d}\hat R(t)}{{\rm d} t}\hat R^\dagger(t) .
\label{div_transf}
\end{align}
Under
the transformation \eqref{div_transf} the Hamiltonian $\hat{\mathcal{H}}(t)$ becomes
\begin{align}
&\hat{\mathcal{H}} (t) =\hat{\mathcal{H}}_x(t) + \hat{\mathcal{H}}_y(t), \label{eq:hamtrans} \\
&\hat{\mathcal{H}}_x(t) = -J \sum_{j,k} \a_{j+1,k}^\dagger \a_{j,k} \ f_x(t; j) + \text{H.c.}, \notag \\
&\hat{\mathcal{H}}_y(t) = -J \sum_{j,k} \a_{j, k+1}^\dagger \a_{j, k} \ f_y(t; j,k) + \text{H.c.}, \notag \\
&
f_x(t; k) =\exp\Big ( {\rm i} \left \{ N \omega t - K_0[ \cos (\omega t + \theta_{k})-\cos(\theta_k)]   \right \} \Big), \notag \\
&
\begin{aligned}
f_y (t; j,k)=&\exp\Big(-{\rm i} \, j K_0 \{ [\cos (\omega t + \theta_{k+1})
\\
&-\cos(\theta_{k+1})] -[\cos (\omega t + \theta_{k})-\cos(\theta_k)]   \} \Big) .
\end{aligned}
\notag
\end{align}

The long-time dynamics associated with the time-periodic Hamiltonian $\hat{\mathcal{H}} (t)$ is well captured by the time-evolution operator over one period of the driving. In the present frame, this operator is expressed as
\be
\hat{\mathcal{U}}(T)\!=\! \mathcal{T} \exp \left ( - {\rm i} \int_0^T\hat{\mathcal{H}} (t) \text{d} t    \right ) \!=\!  \exp \left ( - {\rm i} T \hat{\mathcal{H}}_{\text{F}} \right ),\label{Floquet_operator_moving}
\ee
where $\hat{\mathcal{H}}_{\text{F}}$ denotes the effective (time-independent) Floquet Hamiltonian, and where $\mathcal{T}$ denotes time-ordering. We point out that the dynamics also present a micro-motion, which can be estimated using the method of Refs.~\cite{goldman2014,Goldman:PRA2015,bukov2015,Anisimovas2015}; these effects will not be discussed in the present study, which focuses on the time-averaged dynamics captured by $\hat{\mathcal{H}}_{\text{F}}$.

By inserting the Jacobi-Anger expansion, $$\exp \left ({\rm i} x \cos y \right )\!=\!\sum_{n=-\infty}^{\infty} {\rm i}^n \mathcal{J}_n(x) e^{{\rm i} ny} ,$$ into Eq.~\eqref{eq:hamtrans}, and calculating the time-average of the Hamiltonian $\hat{\mathcal{H}} (t) $ over one period, we obtain a satisfactory approximation for the Floquet Hamiltonian
\begin{align}
&\hat{\mathcal{H}}_{\text{F}}= \hat{\mathcal{H}}_{\rm F}^x+\hat{\mathcal{H}}_{\rm F}^y
\\&\hphantom{\hat{\mathcal{H}}_{\text{F}}}=-J_{\text{eff}}^x \sum_{j,k} \a_{j+1,k}^\dagger \a_{j,k} e^{-{\rm i} N \theta_k} e^{{\rm i} K_0 \cos (\theta_k) } + \mathrm{H.c.}
 \label{floquet} \\
& - \sum_{j,k} J_{\text{eff}}^y (j,k) \a_{j,k+1}^\dagger \a_{j,k} e^{{\rm i} j K_0 \left [  \cos (\theta_{k+1}) - \cos (\theta_k)    \right ]} + \text{H.c.}, \notag
\\
&J_{\text{eff}}^x = J \mathcal{J}_N (K_0), \notag \\
&J_{\text{eff}}^y (j,k) = J \mathcal{J}_0 \left [ 2 j K_0 \sin \left (  \frac{\theta_{k+1} - \theta_{k}}{2} \right )   \right ], \notag
\end{align}
where $\mathcal{J}_{N}$ denotes a Bessel function of the first kind. From expression \eqref{floquet}, one can directly compute the effective magnetic fluxes penetrating the plaquettes of the square lattice. According to the Peierls phase factors in Eq.~\eqref{floquet}, we find the fluxes [Eq.~\eqref{peierls}]
\begin{align}
\Phi (j,k)&\!=\! - N \theta_k  \!+\! K_0 \cos (\theta_k) \!+\! N \theta_{k+1} \!-\! K_0 \cos (\theta_{k+1}) \notag \\
&\quad + (j+1) K_0 \biggl [ \cos (\theta_{k+1}) -\cos (\theta_k) \biggr ] \notag \\
&\quad -j K_0 \biggl [ \cos (\theta_{k+1}) -\cos (\theta_k) \biggr ]  \notag \\
&=N \left ( \theta_{k+1} - \theta_k   \right ), \notag
\end{align}
which in general depend on the location of the plaquettes. In order to reproduce
the Landau gauge (\ref{landau_gauge}), we require
\begin{equation}
\theta_k=  \alpha k.
\end{equation}
Making this choice, we indeed find a uniform flux per plaquette
\begin{equation}
\Phi(j,k)= N \alpha,\label{flux2}
\end{equation}
where $\alpha$ is a parameter that can be tuned in experiments.
A lattice-shaking scheme as proposed by Ref.~\cite{kolovsky},
thus indeed produces a uniform magnetic flux. However, in order to provide
a useful simulation of a uniform magnetic field, the effective mass,
or equivalently the effective  tunneling matrix elements $J_{\text{eff}}^{x,y}$, must
{\em also} be uniform. From Eq.~\eqref{floquet} it can be clearly seen that
while $\Jeff^x$ is constant, $\Jeff^y$ varies with position.
As was pointed out in Ref.~\cite{comment}, this limits
the applicability of this method to a region about the origin where
this variation is sufficiently small.

We have so far restricted our attention to the case of sinusoidal
driving. We can ask if the inhomogeneity in $\Jeff^y$ is a
consequence of this, and thus could be
removed by altering the choice of the periodic driving $f(\omega t)$.
This is, however, not the case. It can
be proven that in any shaking scheme based on Eq.~\eqref{ham_driven},
{\em any} driving function $f(\omega t)$ necessarily produces
a spatially-varying $\Jeff^y$, and thus a non-uniform effective mass.
A general proof of this statement is given in Appendix \ref{appendix:no_unif_tunneling}.

As a final technical remark, we point out that the use of the high-frequency regime
is completely reasonable \cite{goldman2014,Goldman:PRA2015,bukov2015,Anisimovas2015}. Indeed, the perturbative treatment with $\omega$ being the largest energy scale is justified by noting that neither the hopping amplitudes in the ``regauged" \eqref{eq:hamtrans} nor in the original Hamiltonian \eqref{ham_lattice} diverge with $\omega \rightarrow \infty$. This also holds true for the split driving schemes described in the following sections.

\section{Split-driving \label{2_step}}

In order to recover a uniform effective mass, a more complicated form
of driving than that given in Eq.~\ref{ham_driven} must
therefore be used. One such example is the ``split-driving'' scheme
introduced in Ref.~\cite{Creffield:2014}. In this approach, each
period of the driving is split into two steps. In the first, the system
is shaken as in Eq.~\eqref{ham_driven}, but with the tunneling suppressed
along the $y$ direction by suitably changing the optical potential.
In the second step, the $y$ tunneling is restored while the $x$ tunneling
is suppressed. Dividing the driving period in this way means that
it is not subject to
the result proved in Appendix \ref{appendix:no_unif_tunneling}, which assumes that the kinetic-energy term of the Hamiltonian is time independent.

Let us denote the new period of the driving $\tau \!=\! 2 \Delta_t \!=\! 2 M T$, where $M \in \mathbb{Z}$ and $\Delta_t= MT$ is the duration of each
of the individual steps. The time-evolution operator over the period $\tau$ is written as
\be
\hat U (\tau) := \hat U_y \hat U_x , \label{split}
\ee
where $\hat U_x$ describes the evolution of the shaken system of Section \ref{standard_shaking} in the absence of tunneling along the $y$ direction, and where $\hat U_y$ describes normal (undriven) evolution along the $y$ direction only. Based on the results of Section \ref{sine_driving} [Eq.~\eqref{floquet}], we write
\be
\hat U_x := \exp \left (- {\rm i} \Delta_t \hat{\mathcal{H}}_{\text{F}}^x \right ). \label{evol_x}
\ee
Here we have used the fact that $\Delta_t \!=\! T \!\times\! \text{integer}$.
Moreover, we write the bare-tunneling operator $\hat U_y$ as
\begin{align}
&\hat U_y :=\exp \left (- {\rm i} \Delta_t \hat H_y \right ),  \notag \\
&\hat H_y= -J_y \sum_{j,k} \a_{j,k+1}^\dagger \a_{j,k}  + \text{H.c.} \label{evol_y}
\end{align}
Now we can take advantage of the Baker-Campbell-Hausdorff (BCH) formula and approximate the evolution operator over one period $\tau$ in Eq.~\eqref{split} as
\begin{align}
\label{unitary}
\hat U (\tau) &= e^{- {\rm i} \Delta_t \hat H_y} e^{- {\rm i} \Delta_t \hat{\mathcal{H}}_{\text{F}}^x} \simeq e^{- {\rm i} \Delta_t \left (\hat H_y + \hat{\mathcal{H}}_{\text{F}}^x \right) } \\
&= \exp \left (- {\rm i} \tau \hat{\mathcal{H}}_{\text{F}}^{\text{split}} \right ),
\end{align}
where we have introduced a new Floquet effective Hamiltonian
\begin{align}
\hat{\mathcal{H}}_{\text{F}}^{\text{split}}&=-(J_{\text{eff}}^x/2) \sum_{j,k} \a_{j+1,k}^\dagger \a_{j,k} e^{{\rm i} N \theta_k} e^{-{\rm i} K_0 \cos (\theta_k) }  \label{floquet_new} \\
& - (J_y/2) \sum_{j,k} \a_{j,k+1}^\dagger \a_{j,k}  + \text{H.c.} \notag
\end{align}
This shows that the split-driving procedure indeed solves the tunneling-inhomogeneity problem; neither $\Jeff^x$ nor $J_y$ have any position dependence,
and we can set $J_y = \Jeff^x$.
However, contrary to the standard shaking discussed in Section \ref{standard_shaking}, one now faces a new problem: the fluxes are inhomogeneous over the lattice. Indeed, the flux penetrating each plaquette is now given by [see Eq.~\eqref{peierls}]
\begin{align}
\Phi(j,k) &= N \left ( \theta_{k+1} - \theta_k   \right ) - K_0 \left [ \cos (\theta_{k+1}) - \cos (\theta_k)   \right ]. \notag\\
&=N \alpha - K_0 \left[ \cos ((k+1) \alpha ) - \cos (k \alpha)   \right],\label{wrong_flux}
\end{align}
where we again write the phase of the modulation as $\theta_k=  \alpha k$.
Although the flux is not uniform, its
variation is nonetheless bounded, and can be made arbitrarily
small by reducing $K_0$.

\begin{figure}
\begin{center}
\includegraphics[width=0.45\textwidth,clip=true]{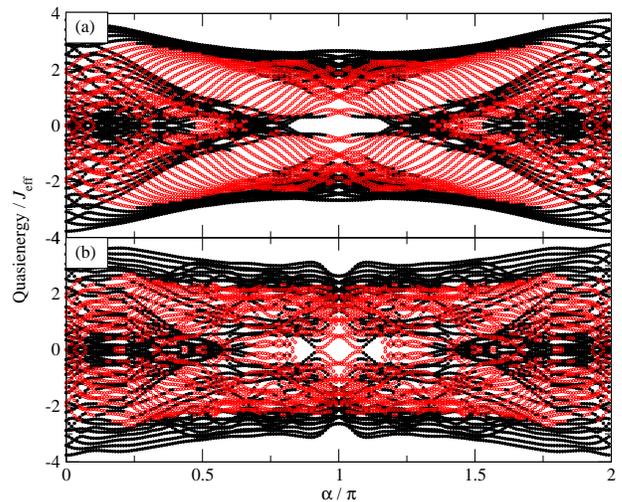}
\end{center}
\caption{Quasienergy spectra for a sinusoidal split-driving.
(a) For weak driving, $K / \omega = 0.2$, the spectrum almost
exactly reproduces the energies of the Harper-Hofstadter Hamiltonian.
(b) For a larger value of the driving,  $K / \omega = 0.8$,
distortions appear in the spectrum.
Parameters of the system:
$8 \times 8$ lattice, $J = 1$, $\omega = 1000 J$.
Red/black symbols indicate edge/bulk states (see text).}
\label{split_spectra}
\end{figure}

In Fig.~\ref{split_spectra} we plot the quasienergies of the
driven system, which are related to the eigenvalues of the
unitary time-evolution operator (\ref{split}) via
$\lambda_n = \exp(-{\rm i} \tau \epsilon_n)$. The quasienergies,
$\epsilon_n$, are the equivalent of the energy eigenvalues
for time-periodic systems, and play an analogous role in determining
the dynamics of the system.
Here, we consider the $N\!=\!1$ resonance, that is, $V_0 \!=\! \omega$, and we apply open boundary conditions in our simulations.
We can see in Fig. \ref{split_spectra}a, that for weak driving,
$K_0 = 0.2$ the quasienergy spectrum is almost identical to
that of the original Harper-Hofstadter Hamiltonian [Eq.~\eqref{tight_binding_harper_hof}] , for this choice of boundary conditions. The red points correspond
to states which have more than $50 \%$ of their weight on the boundary,
and so show the behavior of the edge states, while the black points
show the behavior of the bulk bands. As the lattice is rather
small ($8 \times 8$ sites) these show a smoothened version
of the fractal structure known as the Hofstadter butterfly.
In Fig.~\ref{split_spectra}b we show the results for a
larger value of $K / \omega = 0.8$, and as expected they show a significant
deviation from the Hofstadter result, with the
topological gaps being distorted or destroyed \cite{wilson_loop},
due to the larger spatial
variation of $\Phi(j,k)$ [Eq.~\eqref{wrong_flux}].

Thus while split-driving does yield a uniform effective mass,
its direct application is limited
to small values of $K_0$. The drawback of this
is that a low value of $K_0$
produces a small value of $\Jeff$, meaning that the dynamics of the system
is slow, and that the energy scale of the effective Hamiltonian is
small and thus lower temperatures are required in experiment to
resolve observables of interest  such as the gaps.
We would thus like to find some way to eliminate the space-dependent
term of Eq.~\eqref{wrong_flux}, and thereby avoid this restriction.

\section{Four-step split-driving \label{4_step}}

To remove the inhomogeneity in the flux, one can envisage a simple generalization of the split-driving scheme that will remove the unwanted phase terms
from Eq.~\eqref{wrong_flux}. Suppose that we modify the two-step time-evolution operator $\hat U (\tau)$ in Eq.~\eqref{split} into a \emph{four-step} evolution,
in which each time-step has a duration $\Delta_t$
\be
\hat U (\tau):= \hat U_y \hat U_{\beta} \hat U_x \hat U_{\alpha} . \label{split_four}
\ee
Here $\hat U_{x,y}$ are defined as in Eqs. \eqref{evol_x}-\eqref{evol_y}, and we introduce the operator
\begin{align}
&\hat U_{\alpha}:=\exp \left (- {\rm i} \Delta_t \hat H_{\alpha} \right ) \notag \\
&\hat H_{\alpha}= - \left( K_0 / \Delta_t \right) \sum_{j,k} j  \n_{j,k} \cos (\theta_k)\, .
\label{pulse_pot}
\end{align}
This corresponds to pulsing a linear potential along $x$ during a time $\Delta_t$ while inhibiting tunneling. Note that $\hat U_{\alpha}$ depends both on the $x$- and $y$-coordinate.

Similarly, after applying $\hat U_x$ we introduce the pulse
\begin{align}
&\hat U_{\beta}:=\exp \left (- {\rm i} \Delta_t \hat H_{\beta} \right ),  \notag \\
&\hat H_{\beta}= \left( K_0 / \Delta_t \right)  \sum_{j,k} j \n_{j,k} \cos (\theta_k) = -\hat H_{\alpha}  ,
\end{align}
which corresponds to applying the opposite potential to that in the step $\hat U_{\alpha}$. Altogether, the time-evolution over one period $\tau\!=\!4\Delta_t$ in Eq.~\eqref{split_four} is given by
\begin{align}
\hat U (\tau)&= e^{- {\rm i}\Delta_t \hat H_y} e^{- {\rm i} \Delta_t \hat H_{\beta}} e^{- {\rm i} \Delta_t \hat{\mathcal{H}}_{\text{F}}^x} e^{- {\rm i} \Delta_t \hat H_{\alpha}}  \notag \\
& \simeq e^{- {\rm i} \tau \hat{\mathcal{H}}_{\text{F}}^{\text{4-split}}} \, ,
\end{align}
where we used the BCH formula, and where we have introduced a new effective Hamiltonian for the four-step scheme
\begin{align}
\hat{\mathcal{H}}_{\text{F}}^{\text{4-split}}&=-(J_{\text{eff}}^x/4) \sum_{j,k} \a_{j+1,k}^\dagger \a_{j,k} e^{-{\rm i} N \theta_k}  \label{effective_last} + \text{H.c.} \notag \\
&- (J_y/4) \sum_{j,k} \a_{j,k+1}^\dagger \a_{j,k} + \text{H.c.}
\end{align}
This corresponds to the desired result; a uniform flux {\em and} uniform tunneling rates (i.e.\ a uniform effective mass) over the entire lattice.

\begin{figure}
\begin{center}
\includegraphics[width=0.45\textwidth,clip=true]{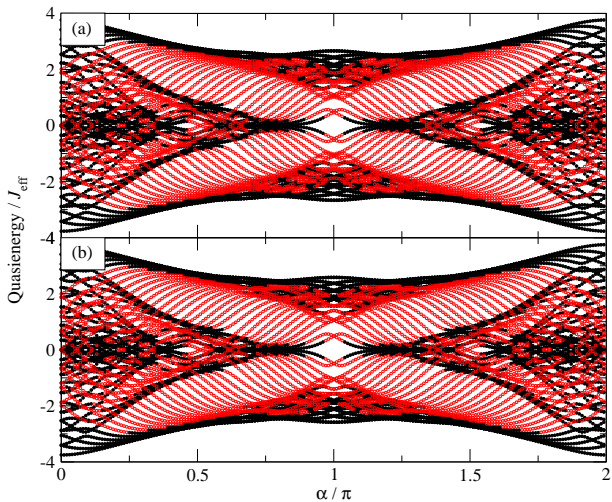}
\end{center}
\caption{Quasienergy spectra for a four-step split-driving
[Eq.~\eqref{split_four}].
(a) For weak driving, $K / \omega = 0.2$, the spectrum
reproduces the Harper-Hofstadter spectrum.
(b) For a very large value of the driving,  $K / \omega = 1.841$ corresponding to
a maximal $\Jeff/J$
the Hofstadter spectrum is again reproduced. This contrasts
with the result in Fig.~\ref{split_spectra}b, where the Hofstadter
structure is lost even for $K / \omega = 0.8$.
Parameters of the system:
$8 \times 8$ lattice, $J = 1$, $\omega = 1000 J$.}
\label{corr_spectra}
\end{figure}

In Fig. \ref{corr_spectra}a we show the quasienergy spectrum for
$K / \omega = 0.2$, and as for the case of two-step split-driving, the results
are practically indistinguishable from the energy spectrum
of the Hofstadter model (a more detailed examination, shown in
Fig. \ref{error}, reveals that the discrepancy is somewhat smaller
than for the case of two-step driving shown in Fig. \ref{split_spectra}a).
In contrast to the two-step
driving however, the agreement with the exact result remains extremely good
as $K / \omega$ is increased. In Fig. \ref{corr_spectra}b we show the
quasienergy spectrum for a large value of $K / \omega = 1.841$, at
which $\Jeff$ takes its maximum value ($\Jeff = 0.582 J$). Again the quasienergies
almost perfectly duplicate the Hofstadter spectrum, confirming that the
correction steps $\hat{U}_{\beta,\alpha}$ indeed completely remove the unwanted phase terms.
This thus gives us the freedom to use whichever value of $K / \omega$
we wish.

We point out that a different four-step driving scheme, which also leads to uniform flux, was proposed by S\o rensen \emph{et al.\ } \cite{sorenson}. In that case, instead of using shaken lattices, the proposal relies on the use of (real) oscillating magnetic fields (see also~\cite{goldman2014,UedaPulse,SpielmanPulse}).

\section{Different waveforms in the two-step approach \label{waveforms}}

The four-step split driving gives the desired result of
completely eliminating the unwanted phase
terms. However it introduces additional complexity into the experimental
realization of the system, and so it is worth considering whether it
is possible to find a means of suppressing these terms within the
two-step approach, by altering the form of the periodic driving
function $f(\omega t)$ in Eq.~\eqref{ham_driven}.

To see how this is possible, we first introduce the function
\be
F(\omega t) := \int_0^t f(\omega t') dt' ,
\label{F_of_t}
\ee
which is related to the velocity of the shaken lattice.
In Appendix \ref{F_factors} we show that
Eq.~\eqref{wrong_flux} generalizes in a straightforward manner
to the case of arbitrary shaking functions $f(\omega t)$,
and that the flux penetrating each plaquette is then given by
\be
 \Phi(j,k) = N \alpha - K_0 \left[ F((k+1) \alpha) - F(k \alpha) \right] \ .
\label{general_phi}
\ee
Ideally we would like to choose $f(\omega t)$ such that $F(k \alpha)$ is
constant, as in that case the flux threading
each plaquette would indeed be uniform, $\Phi(j,k) = N \alpha$.
However, as $f(\omega t)$ is an oscillatory function of time,
$F(\omega t)$ is consequently oscillatory too, and so this condition cannot
be fulfilled except for the trivial case $f(\omega t) = 0$.

Although we cannot therefore achieve the ideal case, we can attempt to
make $F(k \alpha) = 0$ for {\em most} (but not all)
values of $k$. This means that
the synthetic flux will be uniform over large areas of the lattice,
and will only be different for plaquettes which include a hopping phase
(see Eq.~\eqref{hop_phase}) for
which $F(k' \alpha) \neq 0$, that is, plaquettes which have a link
lying along the line $y = k'$ (in units of the lattice spacing).
Our aim is thus to limit the number
of such values of $k'$ to the smallest amount possible.

\begin{figure}
\begin{center}
\includegraphics[width=0.45\textwidth,clip=true]{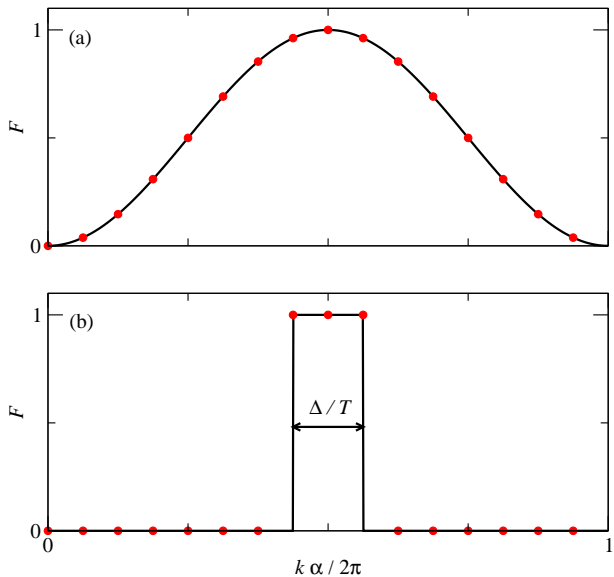}
\end{center}
\caption{The velocity function, $F(k \alpha)$, defined in
Eq.~\eqref{F_of_t}, for two different driving functions.
(a) Sinusoidal driving. The symbols indicate $F(k \alpha)$
for $\alpha = \pi/8$, which give the inhomogeneous component
of the synthetic flux. This spatial dependence produces
the deviations from the Hofstader spectrum seen in
Figs.~\ref{split_spectra}b and c.
(b) Kicked-driving [Eq.~\eqref{kicks}]. In this case $F(k \alpha) = 0$
for most values of $k$, meaning that the synthetic flux is
much more uniform.}
\label{velocity}
\end{figure}

\begin{figure}
\begin{center}
\includegraphics[width=0.45\textwidth,clip=true]{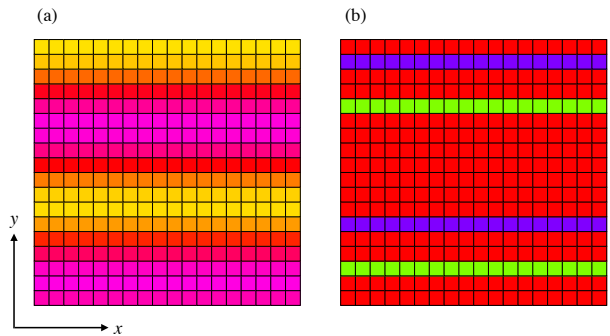}
\end{center}
\caption{Spatial variation of the flux in a
$16 \times 16$ lattice; the color of each plaquette
represents the value of $\Phi(j,k)$ threading it.
a) For a two-step sinusoidal driving, $\Phi(j,k)$ varies periodically
as a function of the $y$-coordinate.
b) For a two-step kicked driving, $\Phi(j,k)$ is generally constant, except
for certain values of $y = k'$ for which $F(k' \alpha)$ is non-zero.
Physical parameters of the system: $K_0 = 0.5, \ \alpha = 0.6$.
Spacing between the kicks, $\Delta = T/8$.}
\label{flux_space}
\end{figure}

We show in Fig.~\ref{velocity}a the behavior of $F(k \alpha)$ for
the case of sinusoidal driving. The symbols indicate values of $k \pi/8$,
that is, the space-dependence of the inhomogeneous component of the flux
when the two-step split-driving is used to simulate a magnetic
flux with $\alpha=\pi/8$. This oscillatory dependence on $k$ produces the
distortions to the Hofstader spectra in Figs.~\ref{split_spectra}b,c
when $K_0$ is not sufficiently small to suppress these terms.
In Fig.~\ref{flux_space}a we show the explicit spatial variation of
the flux arising from this $k$-dependence.

\subsection*{$\delta$-kick configuration}
We now introduce a new scheme utilizing the driving function
defined in the interval $0 \leq t < T$ as
\begin{equation}
f(\omega t) = \delta[t-(T/2 - \Delta/2)] - \delta[t-(T/2 + \Delta/2)] \ .
\label{kicks}
\end{equation}
The full periodic function is obtained by repeating this interval,
producing a sequence of
pairs of $\delta$-kicks separated by a time-interval
$\Delta$.
In Fig.~\ref{velocity}b we show $F(k \alpha)$ for this waveform.
If $\Delta$ is taken to be small, it can clearly be seen
that the $F(k \alpha) = 0$ for the majority of the points. This means
that the synthetic flux will indeed be uniform over large
areas of the optical lattice, with variations only
occurring along certain specific lines $y = k' \alpha$
for which $F(k' \alpha) \neq 0$. The number of these lines
depends on the values of $\Delta$ and $\alpha$, and on their
commensurability (with respect to $T$ and $\pi$). This effect is shown in Fig.~\ref{flux_space}b, where
we show the spatial dependence of the flux-threading each plaquette.
We can clearly see that while the flux per plaquette along certain
lines deviates from the desired value, over large areas of the lattice
the flux is indeed constant.

\begin{figure}
\begin{center}
\includegraphics[width=0.45\textwidth,clip=true]{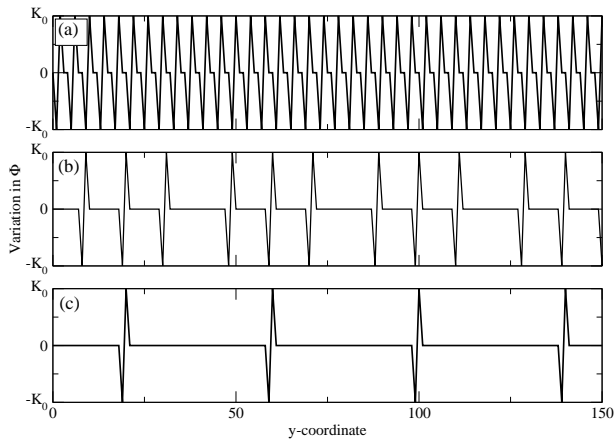}
\end{center}
\caption{ The spatial variation of the flux [see Eq.~(\ref{general_phi})]
for $\delta$-kick driving.
(a) For $\alpha = \pi / 2$ and $\Delta = 0.08 T$, the
flux oscillates rapidly as a function of $y$.
(b) Tuning $\alpha$ away from this value to $\alpha = 0.55 \pi$
substantially reduces the number of oscillations, meaning that the
flux is uniform over longer length scales.
(c)  For $\alpha = 0.55 \pi$, reducing the spacing between
the kicks further (to $\Delta = 0.02 T$) reduces the number of oscillations. The flux
is now constant over length scales of $\sim 50$ lattice
spacings.}
\label{commensurate}
\end{figure}

We show this effect in more detail in Fig.~\ref{commensurate}
where we show the space-dependent flux, i.e. the space-dependent
term in Eq.~\eqref{general_phi}, for
this form of kicked driving for $\alpha = \pi/2$.
This value of $\alpha$ is highly commensurate, and as
a consequence, we can see in Fig.~\ref{commensurate}a
that for a kick-spacing of $\Delta = 0.08 T$, the
flux oscillates rapidly with $y$. This would produce
a highly non-uniform field, although, as with sinusoidal driving,
this inhomogeneity could be controlled by reducing the
size of $K_0$.

In Fig.~\ref{commensurate}b we show the space-dependent flux
for the same value of $\Delta$, but with $\alpha$ now tuned
slightly away from the commensurability condition
to a value of $\alpha = 0.55 \pi$. This slight detuning
has a large effect on the inhomogeneity of the flux, and
we can clearly see that the number of ``bad'' plaquettes
has been considerably reduced. The behavior can be improved
further by reducing the spacing of the $\delta$-kicks. In
Fig.~\ref{commensurate}c we show the results for $\alpha = 0.55 \pi$
and a smaller spacing of $\Delta = 0.02 T$. The flux is
now uniform over distances of $\sim 50$ lattice spacings.
This means that if in experiment the atomic cloud could be
confined to a region of this size, this inhomogeneity would
not be visible; the atomic cloud would then effectively behave according to the standard Harper-Hofstadter model.

\begin{figure}
\begin{center}
\includegraphics[width=0.45\textwidth,clip=true]{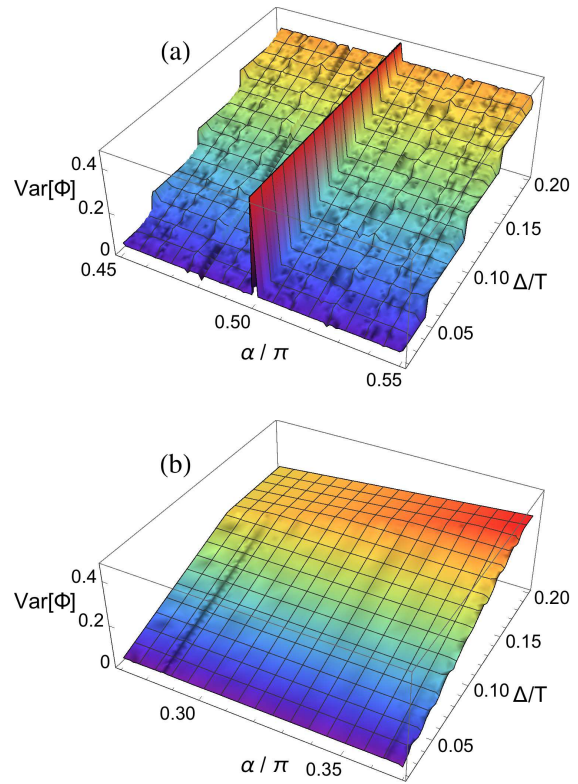}
\end{center}
\caption{The variance of the flux $\Phi(j,k)$ for $\delta$-kick driving,
as a function of $\alpha$ and $\Delta$. For a perfectly uniform field,
the variance will be zero.
(a) For $\alpha = \pi/2$, the performance of the $\delta$-kick
driving is poor. Detuning from this value to reduce the commensurability
substantially reduces the variation in $\Phi(j,k)$, which decreases
as $\Delta$ is reduced.
(b) Away from commensurate values of $\alpha$, the behavior of
the variance is much smoother and falls as $\Delta$ is reduced.}
\label{surface}
\end{figure}

This behavior is summarized in Fig.~\ref{surface}a,
where we plot the variance of the flux as a function
of $\alpha$ and $\delta$. Exactly
at $\alpha= \pi/2$, the $\delta$-kick method performs poorly
for all values of the kick-spacing. Tuning away from this
special value immediately improves its performance, which can be enhanced
further by reducing the size of $\Delta$.

\begin{figure}
\begin{center}
\includegraphics[width=0.45\textwidth,clip=true]{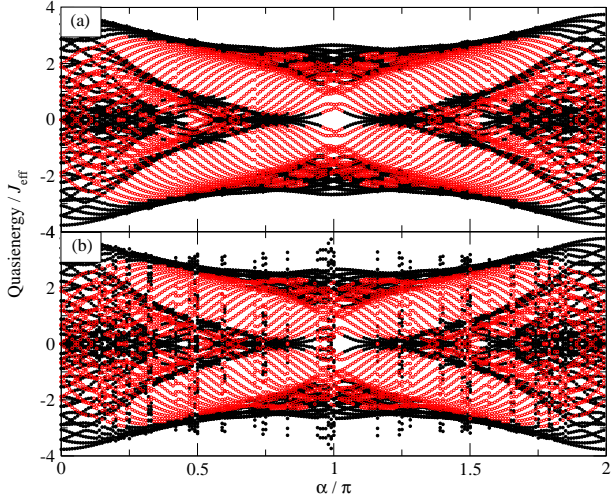}
\end{center}
\caption{Quasienergy spectra for a two-step split-driving
with a kick potential (Eq.~\eqref{kicks}).
(a) For weak driving, $K / \omega = 0.2$, the quasienergy spectrum again
agrees well with the exact Hofstadter result.
(b) For a larger value of the driving,  $K / \omega = \pi$
the main structure of the Hofstadter spectrum is again reproduced.
At certain well-defined values of the flux, however,
the spectrum shows significant deviations.
Parameters of the system: $\Delta = T/32$,
$8 \times 8$ lattice, $J = 1$, $\omega = 1000 J$.}
\label{dirac_spectra}
\end{figure}

For this form of driving, we show in Appendix \ref{effective_delta}
that the effective tunneling is given by
\begin{equation}
\Jeff^x = \frac{2 J}{\pi} \sin \left( \pi \Delta / T \right)
 \sin \left( K / 2 \omega \right) \ ,
\label{renorm_kick}
\end{equation}
where we use the same $N\!=\!1$ resonant condition as we
did for sinusoidal driving.
In Fig.~\ref{dirac_spectra}a we show the quasienergy spectrum
obtained for a low value of the driving strength
$K / \omega = 0.2$, for a kick-spacing of $\Delta = T / 32$.
Clearly the result again agrees very well with the exact
Hofstadter result. In Fig.~\ref{dirac_spectra}b we show
the quasienergies for a much larger value of the driving,
$K / \omega = 3.14$, for which $\Jeff$ takes its maximum
value. Most of the spectrum reproduces the Hofstadter result,
with the exception of certain well-defined values of the
magnetic flux. At these values, the variance
of the flux on the driving parameters resembles that in Fig.~\ref{surface}a.
Elsewhere, however, the variance decays smoothly as $\Delta$ is reduced,
as shown in Fig. \ref{surface}b.
Thus if these specific flux values are avoided,
this form of driving can give excellent performance for a wide range of driving
strengths, as long as $\Delta$ is sufficiently small.

\section{Comparison of methods \label{compare}}

\begin{figure}
\begin{center}
\includegraphics[width=0.45\textwidth,clip=true]{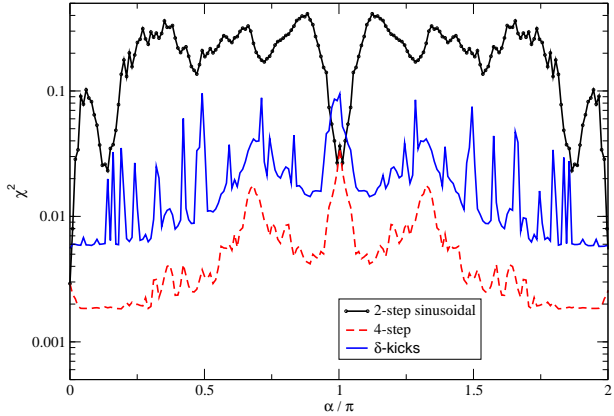}
\end{center}
\caption{$\chi^2$ (see Eq.~\ref{chi2}) for the three different forms of driving for
a low driving strength $K / \omega = 0.2$. The error in the
two-step sinusoidal driving coincides with that of the four-step driving for
$\alpha = 0$ and $\pi$, but elsewhere is notably
higher. The error in the kicked driving varies similarly to that of
the four-step driving, except at certain sharply-defined values of $\alpha$.}
\label{error}
\end{figure}

The performance of the various split-driving schemes can
be made quantitative. We define the figure of merit
\begin{equation}
\chi^2 := \sum_i \left| E^{\mathrm{Hof}}_i - \epsilon_i \right|^2 \ ,
\label{chi2}
\end{equation}
where $\{ E^{\mathrm{Hof}}_i \}$ are the eigenenergies
of the Hofstadter Hamiltonian, and $\{ \epsilon_i \}$
are the quasienergies of the driven system.
We show in Fig.~\ref{error} the behavior of $\chi^2$ for the
three forms of driving, as a function of the magnetic flux.
The error on the sinusoidal driving is the same as for the
4-step driving for $\alpha=0$ and $\pi$, but for other
values of flux the 4-step driving clearly produces results of
higher precision. The precision of the 2-step kicked-driving results in general
mimics that of the 4-step driving, except at the commensurate
values of the magnetic flux  which produce sharp spikes in $\chi^2$.

\begin{figure}
\begin{center}
\includegraphics[width=0.45\textwidth,clip=true]{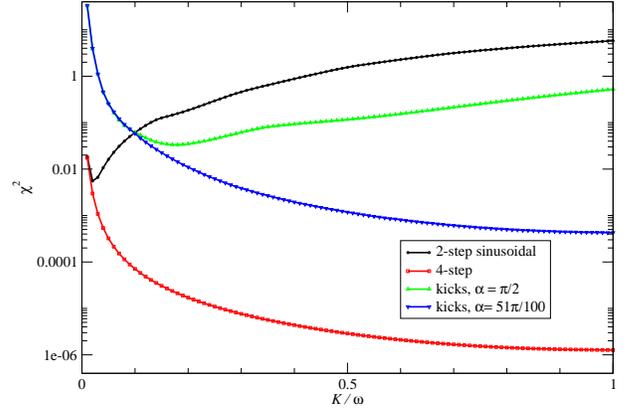}
\end{center}
\caption{$\chi^2$ for the three different forms of driving, for
$\alpha = \pi/2$, as a function of the driving
strength. The error in the four-step driving drops
monotonically with $K / \omega$, while the error in the  two-step split
driving initially falls, then rises. For $\pi/2$ the kicked driving
performs poorly, as this value of $\alpha$
is commensurate with the kick separation. However
for $\alpha = 51 \pi/100$, which avoids the exact
commensurability condition, the error behaves similarly to
that of the four-step driving.}
\label{chi3}
\end{figure}

To further investigate the precision of the methods, we show
in Fig.~\ref{chi3} the $\chi^2$ deviation from the exact
results for a fixed flux of $\pi/2$ as the driving strength
$K / \omega$ is varied. Initially the sinusoidal and the four-step driving
produce results of similar precision, with the deviation dropping
as $K / \omega$ increases.
This behavior is commonly seen in periodically-driven systems; the
static effective Hamiltonian (\ref{floquet}) is obtained as
an approximation in the
high-frequency limit $\omega \gg J$, but the amplitude of the driving
$K$ still remains as another energy scale. When both $K$ and $\omega$ are
large, the time-dependent component of the Hamiltonian completely
dominates the tunneling part, and the quality of the
approximation is enhanced.

The error in the four-step driving is set only
by the error in the Baker-Campbell-Hausdorff decomposition of the original
Hamiltonian and in the high-frequency approximation used to derive
effective Hamiltonians. As we have seen, however, in the two-step
split-driving, the inhomogeneity in the flux grows as $K_0$
increases, making the results diverge from the Hofstadter
spectrum. It is interesting to note that for $\alpha=\pi/2$,
the kicked results behave similarly to those of the
sinusoidal driving. Changing the flux slightly to
a value of $51 \pi/100$, however, tunes the system away from a
commensurablilty condition, and consequently the error falls similarly to
that of the four-step driving.

\section{Shaken lattices, moving lattices, and effective masses \label{general_shaking_section}}

Before concluding, we would like to further discuss how shaken-lattice-based setups compare to other driving schemes, and in particular, how inhomogeneous tunneling matrix elements appear in a more general framework.

Consider a square lattice treated in the single-band tight-binding approximation, described by the static Hamiltonian
\begin{align}
\hat{H}_{0}= \hat T_x + \hat T_y + \Delta \sum_{j,k} j \hat n_{j,k} , \label{statone}
\end{align}
where the nearest-neighbor hopping terms are given by
\begin{align}
&\hat T_x := -J_x \sum_{j,k} \hat a_{j+1,k}^{\dagger} \hat a_{j,k} +  \text{H.c.} ,\notag \\
&\hat T_y := -J_y \sum_{j,k} \hat a_{j,k+1}^{\dagger} \hat a_{j,k} + \text{H.c.} ,\notag
\end{align}
and where $J_{x,y}$ are the hopping matrix elements in the two spatial directions. As discussed in previous Sections, the Hamiltonian in Eq.~\eqref{statone} also includes a constant energy offset $\Delta\!\gg\!J_{x,y}$ between the sites along the $x$ direction.

Now, let us drive this system using a resonant time-modulation with a single harmonic, which we write in the form
\begin{align}
&\hat V (t) = K \sum_{j,k} \hat n_{j,k} \, v (j,k) \, e^{{\rm i} \omega t} + \text{H.c.} ,  \quad \omega = \Delta  , \label{mod_def}
\end{align}
where $K$ is the driving strength, and where we now introduce the general function $v(j,k)$, which describes the spatial dependence of the driving. Then, following a similar analysis to that presented in the previous sections, we find that the system is well described by an effective Hamiltonian of the form~\cite{Goldman:PRA2015}
\begin{align}
\hat{\mathcal{H}}_{\text{eff}} \!=\! \sum_{j,k} \, & \left( \mathcal{J}_x (j,k)e^{ {\rm i} \phi_{j,k}} \hat a_{j+1,k}^{\dagger} \hat a_{j,k} \right. \notag \\
&+ \left. \mathcal{J}_y (j,k) \hat a_{j,k+1}^{\dagger} \hat a_{j,k} \right) + \text{H.c.}, \label{effective_harper}
\end{align}
which corresponds to a hopping Hamiltonian with modified (effective) tunneling amplitudes
\begin{align}
&\mathcal{J}_x (j,k) = J_x \,  \mathcal{J}_1 \left ( 2K_0 \vert \delta_x v(j,k) \vert \right ), \label{amplitudes_strong} \\
&\mathcal{J}_y (j,k) = J_y \,  \mathcal{J}_0 \left ( 2K_0 \vert \delta_y v(j,k) \vert \right ), \qquad K_0=K \omega.\notag
\end{align}
Here $\delta_{x,y}$ denote finite-difference operations along the $x$ and $y$ directions,
\be
\delta_{x} v(j,k)\!=\! v(j+1,k) - v(j,k), \, \, \delta_{y} v(j,k)\!=\! v(j,k+1) - v(j,k),\notag
\ee
and the Peierls phase-factors in Eq.~\eqref{effective_harper} are simply given by $\phi_{j,k}\!=\!\text{arg} [-\delta_{x} v^*(j,k)]$.

Importantly, Eq.~\eqref{amplitudes_strong} indicates how non-uniform effective masses (tunneling amplitudes) appear as a function of the driving function $v(j,k)$. We now illustrate this result below.

\subsubsection{Shaken lattices}
As discussed in this work, shaken lattices are described by a time-modulation of the form \eqref{ham_driven}. For the (single-harmonic) sinusoidal driving considered in Section~\ref{sine_driving}, the spatial function of the drive is given by
\be
v(j,k)= j e^{{\rm i} \theta_k}/2 {\rm i} ,\label{shakenvfun}
\ee
so that the effective tunneling amplitudes are given by [Eq.~\eqref{amplitudes_strong}]
\begin{align}
&\mathcal{J}_x (j,k)= J_x \mathcal{J}_1 (K_0), \notag \\
&\mathcal{J}_y (j,k) = J_y\mathcal{J}_0 \left [ 2 j K_0 \sin \left (  \frac{\theta_{k+1} - \theta_{k}}{2} \right )   \right ], \label{inhom_mass}
\end{align}
as already given in Eq.~\eqref{floquet}. We point out that Eq.~\eqref{amplitudes_strong} directly indicates the fact that the inhomogeneous effective mass in Eq.~\eqref{inhom_mass} directly comes from the inertial force associated with the shaking, which is described by a potential that grows linearly along the $x$ direction, $V(t)\!=\! j m a(t)$, but in which the acceleration $a(t)$ depends on the $k$ coordinate.

\subsubsection{Moving lattices}
In cold atoms, one has the possibility of introducing another type of time-modulation, which is based on ``moving lattices" \cite{bloch,ketterle}. These are potentials that are generated by a single pair of laser beams, with frequency difference $\omega_1 - \omega_2=\omega$, and wave vector difference $\bf{k}_1 - \bf{k}_2=\bf{q}$. In contrast to the shaken lattices discussed above, these ``moving" potentials have the form of a sliding but otherwise fixed potential. Specifically, these moving lattices are described by the driving term
\begin{align}
\hat V (t)&\!=\! 2 K \sum_{j,k} \hat n_{j,k} \cos \left ( \omega t \!+\! q_x j + q_y k  \right )  ,\notag
\end{align}
where, as before, we set the lattice spacing to unity. This corresponds to the on-site energy modulation in Eq.~ \eqref{mod_def} with
\be
v(j,k) = \exp ({\rm i} q_x j) \exp ({\rm i} q_y k).\label{mod_munich_one_bis}
\ee
Importantly, note that the absolute value of this function is trivial and, in particular, it is not linear in the position, which contrasts with the shaken-lattice case in Eq.~\eqref{shakenvfun}. Hence, in this case, the effective tunneling amplitudes are constant and given by [Eq.~\eqref{amplitudes_strong}]
\begin{align}
&\mathcal{J}_x (j,k)= J_x \,  \mathcal{J}_1 \left ( 2\sqrt{2}K_0 \sqrt{1 - \cos q_x} \right ), \label{amplitudes_strong_moving} \\
&\mathcal{J}_y (j,k)= J_y \,  \mathcal{J}_0 \left ( 2\sqrt{2}K_0 \sqrt{1 - \cos q_y} \right ).\notag
\end{align}

This illustrates how two apparently similar approaches (shaken lattices vs moving lattices) can both generate uniform magnetic fluxes, while producing drastically different effective tunneling amplitudes.

\section{Conclusions \label{concs}}

In summary, we have described a series of schemes based on the
periodic shaking of a lattice potential, with the aim of
simulating the physics of a quantum particle
moving on a lattice threaded by a uniform magnetic flux. While simple shaking
produces a uniform flux, the generated effective mass of the particle
varies in space. The split-driving scheme proposed in Ref. \cite{Creffield:2014}
builds on the idea of this form of shaking and
yields a constant effective mass, but the flux produced in this way
has a spatial variation which only becomes negligible for
small shaking amplitudes, 
limiting the application of this method.
By studying the cause of this behavior in detail, we have shown how
to modify the split-driving scheme in order to
obtain the ideal case of uniform flux and constant effective mass.
This can be done by generalizing the two-step split-driving method
to a four-step scheme, which allows the unwanted spatial variation of
the flux to be exactly canceled. Alternatively, within the two-step method,
we have shown that by changing the form of the shaking from the standard
sinusoidal form, the spatial variation of the flux can be substantially reduced
even for large shaking amplitudes. These two methods open the way to realize
the Hofstadter butterfly and Chern bands, and to study phenomena such as the
quantum Hall effect in any experimental situation which permits
shaking of this form. Two prominent examples of this type of system are
ultracold atomic gases held in optical lattices, and photonic crystals.

\bigskip

\acknowledgments
This work has been supported by Spain's MINECO through
Grant No. FIS2013-41716-P. FS acknowledges the support of the
Real Colegio Complutense at Harvard and
the MIT-Harvard Center for Ultracold Atoms.
Research at ICFO has been supported by MINECO (Severo Ochoa grant SEV-2015-0522 and 
FOQUS FIS2013-46768), Catalan AGAUR SGR 874, and Fundaci\'o Cellex. 
NG acknowledges discussions with A. Spracklen and S. Mukherjee; he was supported by the FRS-FNRS Belgium and by the BSPO under PAI Project No. P7/18 DYGEST.

\appendix

\section{Frame transformations \label{appendix:frames}}

The goal of this section is to provide some more context for some of the unitary transformations, such as that in Eq.~\eqref{movingframe}. In particular we will show how to derive the Hamiltonian in Eq.~\eqref{ham_lattice}.
The starting point is the laboratory rest frame in a continuum description.
Our first task is to define
a unitary transformation between a Hamiltonian given in the laboratory frame, and a Hamiltonian where the (accelerated) lattice appears to be at rest.
The authors of Refs. \cite{macgregor2012unitary} and \cite{klink1997} constructed a unitary representation of the Galilean line group, which is the source for the unitary transformations we will use. A concise summary (in first quantization) of the rules of transformation between accelerated frames can be found in Ref. \cite{jdalibard2013}.

Let us start with a many-body Hamiltonian, describing a non-interacting gas subjected to a potential that is shifted arbitrarily in the $x$-direction:
\begin{equation}
\hat{H} = \int_V {\rm d}\vec{r}\,\hat{\Psi}^\dagger(\vec{r})\left[-\frac{\nabla^2}{2m}+V(\vec{r}-d(t)\vec{e}_x)
\right]\hat{\Psi}(\vec{r}) \, ,
\label{eq:contRestframeHam}
\end{equation}
where the field operators fulfill the usual commutation relations
\begin{align}
&\left[\hat{\Psi}(\vec{r}'),\hat{\Psi}(\vec{r})\right]=\left[\hat{\Psi}^\dagger(\vec{r}'),\hat{\Psi}^\dagger(\vec{r})\right]=0~,
\\
&\left[\hat{\Psi}(\vec{r}'),\hat{\Psi}^\dagger(\vec{r})\right]=\delta(\vec{r}'-\vec{r}) \, ,
\end{align}
and $\vec{e}_x$ is the unit vector in the $x$-direction.
The unitary transformation that governs the frame transformation from laboratory to accelerated lattice is defined as
\begin{align}
\begin{aligned}
\hat{U}_{d(t)}:=
&e^{-{\rm i} \frac{m}{2}\dot{d}(t)d(t)\int {\rm d}\vec{r}\, \hat{\Psi}^\dagger (\vec{r})\hat{\Psi}(\vec{r})}
\\
&\hspace{1cm}\times e^{-{\rm i}m \dot{d}(t) \int {\rm d}\vec{r}\, \hat{\Psi}^\dagger(\vec{r}) x\hat{\Psi}(\vec{r})}
\\
&\hspace{2cm} \times e^{{\rm i} d(t) \int {\rm d}\vec{r}\,\hat{\Psi}^\dagger(\vec{r})\frac{1}{{\rm i}}\partial_x\hat{\Psi}(\vec{r})} \, .
\end{aligned}
\label{eq:manyUtrafo}
\end{align}
Note that this transformation operates on both the position
and momentum coordinates of the Hamiltonian, while for example
the momentum shift operator $\hat R(t)$ Eq.~\eqref{movingframe}
acts {\em only} on the momentum.

Transforming the Hamiltonian in Eq.~\eqref{eq:contRestframeHam} results in
\begin{align}
\hat{H}_{\rm accel} &=
{\rm i} \frac{{\rm d} \hat{U}_{d(t)}} {{\rm d}t}\hat{U}^{-1}_{d(t)}+  \hat{U}_{d(t)} \hat{H} \hat{U}^{-1}_{d(t)}
\\
&\begin{aligned}
= \int {\rm d}\vec{r}\, \hat{\Psi}^\dagger(\vec{r}) &\left[-\frac{\nabla^2}{2m}+V(\vec{r})+m \ddot{d}(t)x \right.
\\
&\hspace{1.3cm}\left.
+\frac{m}{2}\ddot{d}(t)d(t) \right]\hat{\Psi}(\vec{r})~.
\end{aligned}
\label{eq:manyBodyHam}
\end{align}
In the tight-binding approximation, this Hamiltonian is equivalent to that in Eq.~\eqref{ham_lattice}.
It is interesting to note that the second factor of Eq.~\eqref{eq:manyUtrafo} is the inverse of the
continuum version of the momentum shift operator $\hat R(t)$ [see Eq. (\ref{movingframe})].
In this spirit we define:
\begin{equation}
\hat R_{\rm c}(t) := e^{{\rm i}m \dot{d}(t) \int {\rm d}\vec{r}\,
	\hat{\Psi}^\dagger(\vec{r}) x\hat{\Psi}(\vec{r})},
\label{R-continuum}
\end{equation}
which is naturally interpreted as the operator that shifts the momenta into
the non-inertial rest frame of the lattice.
This leads to the transformation of $\hat{H}_\mathrm{accel}$ into:
\begin{align}
&{\rm i}\frac{{\rm d}\hat R_{\rm c}(t)}{{\rm d} t}\hat R^\dagger_{\rm c}(t) +\hat R_{\rm c}(t) \hat{H}_{\rm accel}(t) \hat R^\dagger_{\rm c}(t) \\
&\begin{aligned}
= &\int {\rm d}\vec{r}\, \hat{\Psi}^\dagger(\vec{r}) \left[\frac{1}{2m}({\rm i}\nabla+m \dot{d}(t)\vec{e}_x )^2 \right.
\\
&\hspace{2.5cm}\left.
+V(\vec{r})+\frac{m}{2}\ddot{d}(t)d(t) \vphantom{\frac{1}{2m}}\right]\hat{\Psi}(\vec{r})~.
\end{aligned}
\label{trans-Ham-contin}
\end{align}
With the last, space-independent term removed, this is simply the continuum version of the tight-binding Hamiltonian  Eq.~\eqref{eq:hamtrans} in the particular case where $d(t)$ is $y$-independent.

Unlike Eq.~\eqref{eq:manyUtrafo}, which is a full coordinate transformation affecting both positions and momenta, the transformation \eqref{R-continuum} only regauges the momentum. Thus the transformed Hamiltonian \eqref{trans-Ham-contin} and its tight-binding equivalent \eqref{eq:hamtrans} adopt the positions of the accelerated lattice frame while their momenta are those of the lab frame.

\section{Tunneling phases \label{F_factors}}

We consider a general time-periodic shaking function $f(\omega t)$
which enters the Hamiltonian of the system
as written in Eq.~\eqref{ham_driven}.
We now define its antiderivative
\begin{equation}
F(\omega t) = \int_0^t \ f(\omega t') dt' \ .
\end{equation}
Note that we explicitly consider the shaking to be turned on at $t=0$. This
contrasts with many analyses of shaken systems, in which the
shaking is considered to begin at $t \rightarrow - \infty$.
If we now introduce a temporal phase to the shaking, $f(\omega t + \theta)$,
its antiderivative can be explicitly written as
\begin{equation}
\int_0^t f(\omega t' + \theta ) dt' = F(\omega t + \theta) - F(\theta) \ .
\label{antiderivative}
\end{equation}
We can now evaluate the renormalization of the tunneling terms using the
perturbative scheme described in Ref. \cite{cec_prb}. This proceeds
by first calculating the Floquet states of the driven part of the Hamiltonian,
and introducing $H_0$ as a perturbation to them. If we concentrate on a single
link of the lattice, the tunneling from left to right can be evaluated as
\begin{equation}
\Jeff^x / J = \frac{1}{T} \int_0^T dt \exp \left( -{\rm i} V_0 t - {\rm i}
\int_0^t dt' f(\omega t' + \theta) \right) \ ,
\end{equation}
and the tunneling from right to left will simply be the complex conjugate
of this expression.
Considering now the case of a resonant tilt $V_0 = N \omega$, we can
use Eq.~\eqref{antiderivative} to write this in terms of $F$ as
\begin{eqnarray}
\Jeff^x / J &=& \frac{1}{T} \int_0^T dt \exp \left\{ -{\rm i} N \omega t - {\rm i} \left[
 F(\omega t + \theta) -F(\theta) \right] \right\} \nonumber \\
&=&  \frac{1}{T} e^{ {\rm i} F(\theta) }\times \nonumber\\
&\hphantom{=}&\hspace{.3cm} \int_0^T dt \exp \left\{ -{\rm i} N \omega t - {\rm i} F(\omega t + \theta) \right\} .
\label{renorm}
\end{eqnarray}

Since $f(\omega t)$ is a $T$-periodic function of time, $F(\omega t)$ is
also $T$-periodic, and so is $\exp[{\rm i} F(\omega t)]$. As a result
it can be expanded in a Fourier series as
\begin{equation}
e^{ {\rm i} F(\omega t) } = \sum_{m=-\infty}^{\infty} \gamma_m
e^{ {\rm i} m \omega t } \ .
\end{equation}
Note that for the specific case of sinusoidal driving, this Fourier
series is exactly the Jacobi-Anger expansion.
Substituting this expression in Eq.~\eqref{renorm} gives the result
\begin{equation}
\Jeff^x / J = \frac{1}{T} e^{ {\rm i} F(\theta) }
\int_0^T dt \exp \left( -{\rm i} N \omega t \right)
 \sum_{m} \gamma_m e^{{\rm i} m (\omega t + \theta )} \ .
\end{equation}
Exchanging the order of the integration and the summation gives the
final result
\begin{equation}
\Jeff^x / J =  \gamma_N e^{ {\rm i} ( N \theta + F(\theta) )} \ ,
\label{hop_phase}
\end{equation}
and thus we can see that the amplitude of the $x$-hopping is reduced by a
factor of $\gamma_N$, and that it acquires a phase of $N \theta + F(\theta)$.
For sinusoidal shaking the renormalized amplitude is simply
$\gamma_N = {\mathcal J}_N(K_0)$, and the acquired phase
is $N \theta - K_0 \cos \theta$, in agreement with
the result derived previously in Eqs.\eqref{floquet} and \eqref{wrong_flux}.

\section{Impossibility of uniform tunneling \label{appendix:no_unif_tunneling}}

In this section we will prove that a non-zero uniform magnetic flux
is incompatible with uniform effective hopping, when considering
a general periodic driving of the type given in Eq.~\eqref{ham_driven}
\begin{equation}
\hat{H}_{\rm latt}(t)= \hat{H}_0+\sum j D(t,k) \hat{n}_{j,k}~.
\label{eq:gendrivham}
\end{equation}
Adjusting the definition of the unitary transformation in Eq.~\eqref{R-trans-TB} leads to
\begin{equation}
\hat{R}(t) = \exp\left({\rm i} \int_{0}^t \sum_{j,k} j D(t',k)\hat{n}_{j,k} {\rm d} t' \right) \ .
\end{equation}
The transformed Hamiltonian then becomes
\begin{align}
\begin{aligned}
\hat{H}'(t) = -J \sum_{j,k}& f_x'(t;k) \hat{a}^\dagger_{j+1,k}\hat{a}_{j,k} + {\rm H.c.} \\
- & J \sum_{j,k} f_y'(t;j,k) \hat{a}^\dagger_{j,k+1}\hat{a}_{j,k} + {\rm H.c.}
\end{aligned}~
\end{align}
where now
\begin{align}
 & f_x'(t;k) = e^{{\rm i} \int_{0}^t D(t',k) {\rm d} t'}\\
 & f_y'(t;j,k)= e^{{\rm i} \int_{0}^t j\left[D(t',k+1)-D(t',k)\right]{\rm d} t'} \, .
\end{align}
Again we define the effective Hamiltonian $H'_{\rm eff} = \frac{1}{T} \int_0^T \hat{H}'(t) {\rm d} t$, taking advantage of the periodicity of the Hamiltonian
$\hat{H}'(t+T)=\hat{H}'(t)$.
We will now show the following: Uniformity of the magnitude of the effective hopping implies that
the magnetic flux is zero everywhere.
A very important assumption for this statement to be true is that the
absolute value of the hopping matrix elements of the undriven Hamiltonian is independent of position and time.
Such hypotheses are implicit in the fact that $\hat{H}_0$ in Eq. \eqref{eq:gendrivham} is
a static tight-binding Hamiltonian with constant tunneling matrix elements $J$ [see Eq.~\eqref{tight_binding}].

Since the problem in Section \ref{sine_driving} and in Ref. \cite{comment} was the inhomogeneity
of the hopping matrix elements in the $y$ direction, we focus on that question.
So let us consider the squared modulus of the time-averaged (possibly complex), dimensionless effective hopping amplitude:
\begin{align}
C &= \left|J^y_{\rm eff}(j,k)\right|^2/J^2 \\
&= \frac{1}{T^2}\int^T_0\int^T_0 e^{{\rm i} j [G(t,k)-G(t',k)]}{\rm d} t {\rm d} t'~,
\label{eq:consteffhopp}
\end{align}
where $J$ is the (constant and uniform) magnitude of the hopping energy in the undriven Hamiltonian and
\begin{equation}
G(t,k) = \int_{0}^t [D(t',k+1)-D(t',k)]{\rm d} t'~.
\end{equation}
We wish to prove that, if $C$ as defined in \eqref{eq:consteffhopp} is independent of $j,k$, then the magnetic flux of the time-averaged driven Hamiltonian must be zero.

If Eq.~\eqref{eq:consteffhopp} has to hold for all $j$ with $C$ constant, then in particular it must hold
for $j=0$; therefore $C=1$. If we set
$j = 2 q~, q \in \mathbb{Z}\backslash_0$, then Eq.~\eqref{eq:consteffhopp} implies
\begin{align}
 &\hspace{-1cm}
\begin{aligned}
1=\frac{1}{2 T^2}\int^T_0 \int^T_0& \left(  e^{{\rm i} 2q[G(t,k)-G(t',k)]}
\right.\\
&\hspace{.2cm}\left.
+e^{-{\rm i} 2q[G(t,k)-G(t',k)]}\right) {\rm d} t {\rm d} t'
\end{aligned}
\\
 &= \frac{1}{T^2}\int^T_0 \int^T_0 \cos \Big( 2q[G(t,k)-G(t',k)]\Big){\rm d} t {\rm d} t' \nonumber
\\
&
\begin{aligned}
 = \frac{1}{T^2}&\int^T_0 \int^T_0 \times \\
 &\bigg[1- 2 \sin^2 \Big(q[G(t,k)-G(t',k)]\Big) \bigg] {\rm d} t {\rm d} t' \ ,
\end{aligned}
\nonumber
\end{align}
and thus
\be
 0 = \int^T_0 \int^T_0  \sin^2 \Big(q[G(t,k)-G(t',k)]\Big) {\rm d} t {\rm d} t'~.
\ee
Since the integrand in the above is positive the last equation can only be satisfied when
\begin{equation}
G(t,k)-G(t',k)=\pi l(k)~,
\end{equation}
where $l(k)\in \mathbb{Z}$. Differentiating the expression with respect to
$t$ results in
\begin{equation}
D(t,k+1)-D(t,k) = 0 \, .
\end{equation}
We thus conclude that $D(t,k)=D(t)$ has to be independent of $k$. This however implies that
all the Peierls phases vanish and hence the magnetic flux is trivially zero.
In conclusion, we have shown that a uniform mass implies a zero flux, and thus
a non-zero flux is incompatible with a uniform mass in this simple but general shaking scheme.

\section{$\delta$-kick driving \label{effective_delta}}

Using the same approach as above, we now analyze the $\delta$-kick
shaking, introduced in Eq.~\eqref{kicks}, to extract the renormalization
of the amplitude of the tunneling.

In the interval $0 \leq t < T$, the shaking function is given
by
\begin{equation}
f(t) = \delta(t-T_1) - \delta(t-T_1 - \Delta) \ ,
\end{equation}
that is, a pair of $\delta$-kicks separated by a time interval of $\Delta$.
Its antiderivative, $F$, is then given by:
$$
F(t) =
\begin{cases}
0, \ 0 \leq T_1 \\
1, \ T_1 \leq t < T_1 + \Delta \\
0, \ T_1 + \Delta \leq t < T \ .
\end{cases}
$$
An example of this function is plotted in Fig.~\ref{velocity}b.

We can now evaluate the effective tunneling
\begin{equation}
\Jeff^x / J = \frac{1}{T} \int_0^T dt \exp \left[ -{\rm i} N \omega t + (K / \omega) F(t)
\right] \ .
\end{equation}
This integration is straightforward to evaluate, leading to the result
\begin{equation}
| \Jeff^x / J | = \frac{2}{N \pi} \sin \left( N \omega \Delta / 2 \right)
\sin \left( K / 2 \omega \right) \ .
\end{equation}
Eq.~\eqref{renorm_kick} is then recovered for the specific resonant case
of $N = 1$.

\end{document}